\documentclass{article}

\usepackage{arxiv}

\usepackage[utf8]{inputenc} % allow utf-8 input
\usepackage[T1]{fontenc}    % use 8-bit T1 fonts
\usepackage{hyperref}       % hyperlinks
\usepackage{url}            % simple URL typesetting
\usepackage{booktabs}       % professional-quality tables
\usepackage{amsfonts}       % blackboard math symbols
\usepackage{nicefrac}       % compact symbols for 1/2, etc.
\usepackage{microtype}      % microtypography
\usepackage{lipsum}
\usepackage{graphicx}

\usepackage{layouts}
\usepackage{graphicx}
\usepackage{color}
\usepackage{url} %this package should fix any errors with URLs in refs.
\usepackage{amsmath,amssymb}
\usepackage{epsf} 
\usepackage{psfrag}
\usepackage{graphicx}
\usepackage{adjustbox}
\usepackage{cancel}
\usepackage{lineno}
\usepackage{siunitx}
\usepackage{textcomp}
\newcommand{\pt}{\partial}
\newcommand{\pderiv}[2]{\frac{\pt #1}{\pt #2}}

\begin{document}

\title{Investigations into the opening of fractures during hydraulic testing using a hybrid-dimensional flow formulation%\thanks{Grants or other notes
}


\author{
Patrick Schmidt \\
             University of Stuttgart, \\
              Institute of Applied Mechanics (CE), \\
              Pfaffenwaldring 7, D-70\,569 Stuttgart, \\
              Germany \\
              \texttt{patrick.schmidt@mechbau.uni-stuttgart.de}           %  \\
           \And
           Holger Steeb,  \\
             University of Stuttgart, \\
              Institute of Applied Mechanics (CE), \\
              Pfaffenwaldring 7, D-70\,569 Stuttgart, \\
              Germany \\
              \texttt{holger.steeb@mechbau.uni-stuttgart.de}
              \And
           J\"org Renner, \\
             Ruhr-Univerit{\"a}t Bochum, \\
              Institute of Geology, Mineralogy and Geophysics, \\
              D-44\,801 Bochum, \\
              Germany \\
              \texttt{joerg.renner@ruhr-uni-bochum.de}
}

\maketitle

\begin{abstract}
We applied a hybrid-dimensional flow model to pressure transients recorded during pumping experiments conducted at the Reiche Zeche underground research laboratory to study the normal opening behavior of fractures due to fluid injection. Two distinct types of pressure responses to flow-rate steps were identified and numerically modelled using a radial-symmetric flow formulation for a fracture that comprises a non-linear constitutive relation for the contact mechanics governing reversible fracture surface interaction. These two groups represent radial-symmetric and plane-axisymmetric flow regimes from a conventional pressure-diffusion perspective. A comprehensive parameter study into the sensitivity of the applied hydro-mechanical model to changes in characteristic fracture parameters revealed an interrelation between fracture length and normal fracture stiffness that yield a match between field observations and numerical results. Fracture stiffness values increase with corresponding fracture length. Decomposition of the acting normal stresses into a stresses associated with the deformation state of the global fracture geometry and the contact stresses indicates that geometrically induced stresses contribute the more the lower the total effective normal stress and the shorter the fracture. Separating the contributions of the local contact mechanics and the overall fracture geometry to fracture normal stiffness indicates that the latter, the geometrical stiffness, constitutes a lower bound for total stiffness; its relevance increases with decreasing fracture length, too. Our study demonstrates that non-linear hydro-mechanical coupling can lead to vastly different hydraulic responses and thus provides an alternative to conventional pressure-diffusion analysis that requires changes in flow regime to cover the full range of observations.

\keywords{Hydro-mechanical fracture flow \and Hybrid-dimensional modeling \and Fracture contact mechanics \and Hydraulic testing of fractures \and Reiche Zeche underground research laboratory }
\end{abstract}

\section{Introduction}
Estimation of a reservoir's effective hydraulic properties requires a consistent analysis of experimentally determined pressure and flow transients \cite{muskat1938flow,fetter2001}. For individual fractures, simple analytical models for pressure diffusion have been applied when their intersection with boreholes classified them as axial or radial \cite{matthews1961,matthews1967,horne1995,bourdet1989}. Analytical models based on solutions of the diffusion equation for fluid pressure for constant flow-rate steps document distinct differences in pressure response for the one-dimensional and the radial flow associated with axial and radial fractures, respectively. Rocks with a dense array of randomly oriented fractures may justify their treatment as porous media, leading to radial flow, too. Mathematically, the full range of responses can be addressed by regarding the dimension of the flow to be a parameter \cite{barker1988}. However, hydro-mechanical phenomena, such as reverse water-level fluctuations in distant monitoring wells \cite{rodrigues1983,kim1997} or insensitivity of pressure responses to increases in flow-rates, so called jacking \cite{quirion2010} cannot be reproduced by pressure diffusion models and result in inaccurate approximation of the effective fracture characteristics \cite{vinci2014,murdoch2006,cappa2018}. Despite the growing number of treatments of hydro-mechanical coupling  \cite{murdoch2006,girault2015,girault2016,castelletto2015,schmidt2019}, the understanding of the influence of basic geometrical and mechanical properties of fractures on their hydraulic response to flow-rate or pressure perturbations appears still limited. 

Non-Local fracture deformations triggered by perturbations of the fluid pressure along a fracture induce changes in permeability and volume of fractures with a direct impact on flow and storage characteristics and therefore on how the perturbations evolve with time and spread in space \cite{vinci2014II,vinci2014,quintal2016,berre2019}. Accounting for hydro-mechanical interaction throughout numerical fitting of pressure $p$ and flow-rate $Q$ transients is a non-trivial task and requires consistent evaluation of the balance equations in an efficient manner. Evaluation of fracture opening or closing in response to a perturbation of the equilibrium state requires to consider the acting normal stresses owing to their control on the mechanical interaction between the fracture surfaces in contact. For example, large numbers of single Hertzian contacts have been invoked to characterize the mechanical interaction of two mated fracture surfaces \cite{timoschenko1987,greenwood1966,cook1992}. Responses of these contacts to changes in shear and normal stress result in changes of the effective fracture aperture \cite{goodman1976,bandis1983}. 
Fracture opening does not depend on local contact mechanics alone but also on the geometrical stiffness of the fracture \cite{murdoch2006}. Despite the importance of fracture stiffness for the interpretation of pumping operations, little work has been devoted to decompose these two contributions. Here, we analyze pressure transients from pumping tests conducted at the Reiche Zeche underground research laboratory, where the injection borehole penetrates a fractured rock. From a classical pressure-diffusion perspective, the hydraulic responses of the tested intervals mimic that of either radial-symmetric (positive-tangent group) or axial-symmetric (pressure-plateau group) fractures. Yet, logging and impression-packer results do not support this simple association of fracture geometry and hydraulic response. We employ a hydro-mechanical model considering radial-symmetric conditions, as applying for a radial fracture following a monolithic numerical implementation \cite{schmidt2019} to study the origin of the distinct pressure transients. Specifically, we studied the sensitivity of the hydro-mechanical model to the variation of characteristic fracture properties to identify best fits to the field data and the interdependence between the characteristic fracture properties. We highlight the influence of the two contributions to normal stiffness on the opening behaviour by separating the opening controlled by geometrical and normal contact stiffness as a function of the applied fluid pressure. We demonstrate that -with appropriate sets of parameters- distinctly different pressure responses can in principle be explained by a single model for hydro-mechanical effects, which contrasts the necessity to advocate differences in fracture orientation for pressure diffusion models.

\section{Test site and experimental approach}
\subsection{The Reiche Zeche underground research laboratory}
As part of the research program of STIMTEC, a cooperative project investigating the creation and growth of fractures in crystalline rocks to develop and optimise hydraulic STIMulation TEChniques \cite{dresen2019,rennerNews} we performed hydraulic tests in the research mine Reiche Zeche (Rich Mine), Freiberg (Germany). The average overburden at the test site amounts to about 130 m. The foliation of the fine- to medium-grained biotite gneiss dips 5 to 15° in south-east-direction. The gneiss is penetrated by fairly randomly oriented joints with an average separation of several decimeters. Fracture counting on retrieved cores yield 4.4$\pm$2.5, but intact sections with a length of 1 to 2 m occur. In the test volume of about 40 m $\times$ 50 m $\times$ 20 m, two to three steeply dipping, east-west trending damage zones were identified with a variable width between decimeters and a few meters. 

The injection borehole BH10 with a length of 63 m and a radius of 0.038 m has a strike of N31°E and a dip of 15° from the horizontal and thus the borehole axis intersects the foliation at an angle of 20 to 30°. Ultrasonic transmission of the test volume as well as laboratory experiments on cores revealed a pronounced anisotropy in elastic parameters. Ultrasonic waves travel almost two times faster in the direction of the foliation than perpendicular to it. Dynamic and static Young’s moduli in the two directions differ by 10 to 15 \%, with the low modulus observed for loading perpendicular to the foliation \cite{adero2020}.

\subsection{Experimental procedure}
Experiments were performed with a double-packer probe of Solexperts GmbH, Bochum, Germany, consisting of two inflatable packers isolating an injection interval of about 0.7 m length. The probe is equipped with uphole and downhole pressure gauges, and an uphole flowmeter, all sampled with 0.2 s. Flowrates measured uphole, i.e., outside of the borehole at the pump, were corrected for the storage capacity of the injection system to derive the true flow into the rock. The storage capacity was determined in a calibration experiment, for which the probe was inserted in a hollow steel cylinder. 

The uniformly applied pumping protocol comprised a sequence consisting of a) injection (with rates of 2–10 l/min) until breakdown pressure was reached, the fracking, and subsequent shut-in phase, b) three repeat injections, the refracs, with moderate rates of 3 l/min at maximum, each again followed by a shut-in phase, and c) step-rate tests involving several phases of injection with constant flow rates, successively increased from below 1 l/min to about 5 l/min. The pressure response in these step-rate tests constrains the jacking pressure, the fluid pressure at which the fracture(s) intersecting the borehole wall open. Opening is indicated by a significant increase in injectivity, the ratio between flow-rate and pressure. Impression-packer tests were performed after the entire pumping sequence to document fracture traces on the borehole wall.   

\subsection{Intervals and selected data sets}
The data used here represent part of the results of the step-rate tests performed in six intervals at depths of 24.6 m, 40.6 m, 49.7 m, 51.6 m, 55.7 m, and 56.5 m. Logging before and after fluid injection with an acoustic televiewer and impression-packer tests revealed evidence for pre-existing and induced fractures with a range of orientations (Table \ref{tab:interval_props}). We consider the circumferential fracture traces to represent radial fractures even though they do not intersect the borehole axis at a right angle as strictly required. Also, the traces classified as "axial" do not match this end-member geometry in a strict sense but the tilt to the borehole axis typical of en-echelon hydro-fractures occurring in boreholes that do not follow a principal stress axis \cite{zoback2007}. Actually, the short traces of interval 51.6 m are not well constrained at all. Furthermore, it is impossible to associate the observed pressure transients with a specific fracture trace when intervals exhibit multiple traces. This situation is not unusual but representative of what an interpreter typically faces when tests are performed in crystalline rocks.

\renewcommand{\arraystretch}{1.5}
\begin{table*}[htb]
\label{tab:interval_props}
\caption{Interval characteristics }
%\centering
\begin{adjustbox}{max width=\textwidth}
\begin{tabular}{llrrl}\hline
depth & label$^\dagger$ & $t^\mathrm{p}_{1/3}$$^\ddagger$ & $t^\mathrm{si}_{1/2}$$^\intercal$ & orientation of fracture traces\\
(m) & & (s) & (s) &  \\ \hline
24.6 & $M^t_a$ & $\gtrsim$433 & $\gg$226 & 1 parabola induced\\
40.6 & $M^t_b$ & 10 & $>$170 &  1 pre-existing circumferential, 2 axial\\
49.7 & $M^p_b$ & 24 & 8 &  1 pre-existing circumferential, 2 pre-existing parabola, 1 pair axial induced\\
51.6 & $M^p_a$ & 7 & 3 &  several short axial\\
55.7 & $M^t_c$ & 6 & 28 &  1 pre-existing circumferential, 1 pair axial induced \\
56.5 & $M^p_c$ & 3 & 33 &  1 pre-existing circumferential, 1 axial induced, \\
\hline
\end{tabular}
\end{adjustbox}
$\dagger$ classification of data set (see Figure \ref{fig_res:exp_data})\\
$\ddagger$ time it took for a pressure pulse to decay by $1/3$ before the stimulation phase\\
$\intercal$ time it took for pressure to decay by $1/2$ during the shut-in phase after the step-rate test
\end{table*}
\renewcommand{\arraystretch}{1.0}

\begin{figure}[htb]
    \resizebox{1.0\textwidth}{!}{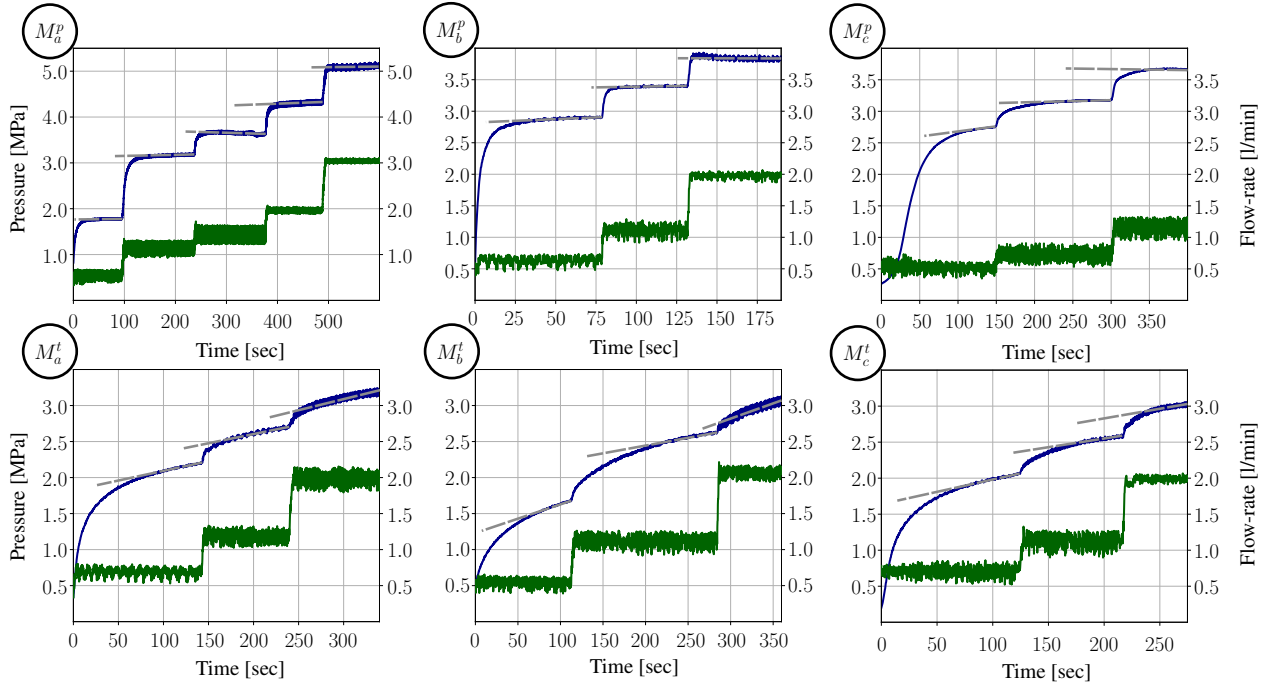}
    \caption{The step-rate test data selected from the six intervals is divided in the pressure-plateau group labeled with $M^p$ and the positive-tangent group labeled with $M^t$. The dark blue lines are associated to the recorded pressure data, the dark green lines show the step-wise increasing flow-rate records and the dotted grey lines represent tangents to the pressure transient at the end of a flow-rate step.}
    \label{fig_res:exp_data}
\end{figure}

We selected three to five of the first low flow-rate steps for the six intervals, addressed as data sets $M^p_{a}$ to $M^p_{c}$ and $M^t_{a}$ to $M^t_{c}$ (Table \ref{tab:interval_props}, Figure \ref{fig_res:exp_data}). The selection aimed to restrict to pressure and flow-rate couples, for which the proposed elastic model most likely applies. For some intervals, seismic activity was observed and therefore its absence during flow-rate steps could used as criterion for "elastic" response. The pressure transients induced by the step-wise increase of flow rate differ for the six intervals. We distinguish two groups of pressure evolution during a step. Pressure responses with flat tangents are evident in data sets $M^p_{a}$ to $M^p_{c}$, a subset of our data that we will address as "pressure-plateau group". In contrast, the data sets $M^t_{a}$ to $M^t_{c}$ exhibit continuously increasing pressure, the "positive-tangents group". For either group, however, the sensitivity of pressure level to flow rate diminishes with increasing flow rate, the observation interpreted as jacking.

\section{Numerical Method}
When the aim is characterization rather than modification, hydraulic testing of fractures is performed below critical pressures for fracture extension, e.g., indicated by a decrease in injection pressure during constant-rate pumping, when rapid addressed as breakdown often related to tensile hydro-fracturing or when occurring over extended time periods or in a succession of small drops probably related to shearing events, either possibly accompanied by characteristic seismic activity. For tests performed at moderate injection pressures, fracture length can thus be treated constant and effects of changes in shear stress and therefore shear stiffness can be neglected. Fractures induced by hydraulic fracturing are expected to be oriented normal to the direction of the least principal (compressive) stress \cite{murdoch2006} so that they intrinsically fulfill the assumption of negligible shear stress. In the context of hydraulic characterization of fractures, their contact mechanics may thus be reduced to an account of their normal stiffness \cite{pyraknolte2000}. Hence, we apply a hydro-mechanical model considering the constitutive relation of normal fracture surface contact following a monolithic numerical implementation \cite{schmidt2019} for the numerical fitting of characteristic fracture parameters, initial width/aperture $\delta_0$, length $l_\text{Fr}$, and normal stiffness parameter of a fracture $E_\text{Fr}$. Its central aspects are recapitulated in the following, before details of the performed parameter search are presented.

\subsection{Governing Equations}
Flow processes of weakly-compressible, viscous fluids in high-aspect ratio fractures motivate the assumption of creeping flow conditions between two locally parallel plates, for which the balance of momentum reduces to a Poiseuille-type formulation \cite{witherspoon1980,vinci2014a}, i.e., the relative fluid velocity $\mathbf{w}_\mathfrak{f}$ is proportional to the pressure gradient $\text{grad}\,p$. In our continuum description, the associated cubic law
\begin{linenomath*}
\begin{equation}
\label{eq:hda_pressure_flux}
\mathbf{w}_\mathfrak{f} = -\frac{\delta(\mathbf{u})^2}{12\,\eta^{\mathfrak{f}R}} \, \text{grad} \, p
=: -\frac{k^\mathfrak{s}_{Fr}}{\eta^{\mathfrak{f}R}} \, \text{grad} \, p
\end{equation}
\end{linenomath*}
is locally evaluated in the fracture domain $\Gamma^{Fr}$, i.e., on the level of a material point $\mathcal{P}(\mathbf{x},\,t)$, where $\mathbf{x}$ denotes its position vector, $\mathbf{u}{(\mathbf{x},\,t)} = \mathbf{x} - \mathbf{X}$ the fracture deformation relative to the reference position vector $\mathbf{X}$, $t$ time, and $\eta^{\mathfrak{f}R}$ the dynamic fluid viscosity. The locally evaluated, deformation-dependent permeability is identified as $k^\mathfrak{s}_{Fr}(\mathbf{x},\,t) = \delta^2 / 12$ considering $\delta(\mathbf{u}{(\mathbf{x},\,t)})$ to be the effective fracture aperture.

The hybrid-dimensional formulation is obtained by inserting the balance of momentum into the balance of mass, derived for a deformable fracture. The outcome of a dimensional analysis of the resulting partial differential equations suggests that quadratic and convective terms are negligible \cite{vinci2014a,vinci2014b,vinci2015} and the accordingly reduced hydro-mechanical governing equation reads
\begin{linenomath*}
\begin{equation}
\label{eq:governing}
\underbrace{\vphantom{\frac{1}{12 \,\eta^{\mathfrak{f}R}\, \beta^{\mathfrak{f}}} \pderiv{}{x}\left(\delta^2 \pderiv{p}{x}\right)} 
\pderiv{p}{t}}_{\MakeUppercase{\romannumeral 1{)}}} - \underbrace{\frac{\delta^2}{12\,\, \eta^{\mathfrak{f}R} \beta^{\mathfrak{f}}} \text{div} \left( \text{grad} \, p\right)}_{\MakeUppercase{\romannumeral 2{)}}} + \underbrace{\vphantom{\frac{1}{12\,r\, \eta^{\mathfrak{f}R}\, \beta^{\mathfrak{f}}} \pderiv{}{r}\left(\frac{\delta^2}{r} \pderiv{p}{x}\right)} \frac{1}{\delta\, \beta^{\mathfrak{f}}}\pderiv{\delta}{t}}_{\MakeUppercase{\romannumeral 3{)}}}= \underbrace{\vphantom{\frac{1}{12\,r\, \eta^{\mathfrak{f}R}\, \beta^{\mathfrak{f}}} \pderiv{}{r}\left(\frac{\delta^2}{r} \pderiv{p}{x}\right)} \frac{q_{lk}}{\delta\, \beta^{\mathfrak{f}}}}_{\MakeUppercase{\romannumeral 4{)}}},
\end{equation}
\end{linenomath*}
comprising a transient $\MakeUppercase{\romannumeral 1{)}}$, a diffusion $\MakeUppercase{\romannumeral 2{)}}$, a coupling $\MakeUppercase{\romannumeral 3{)}}$, and a leak-off term $\MakeUppercase{\romannumeral 4{)}}$, where $\beta^{\mathfrak{f}}$ denotes the fluid compressibility, and $q_{lk}$ leak-off, i.e., the flow-rate from the fracture into the surrounding rock mass. The deformation dependent effective fracture aperture $\delta(\mathbf{u}(\mathbf{x},t))$ contributes to the characteristic diffusion process by term $\MakeUppercase{\romannumeral 2{)}}$ and to volume changes of the fluid domain by term $\MakeUppercase{\romannumeral 3{)}}$, which strongly couples the solution of the fracture-flow domain to the deformation state of the surrounding matrix. 

The rock matrix surrounding the fracture might be treated by purely elastic or by biphasic poro-elastic (e.g., Biot's theory \cite{biot1941}) formulations depending on the application in mind. For the typically substantial difference in the characteristic times of pressure diffusion in the fracture and in a surrounding crystalline rock, a biphasic description results in oscillations of the pore-pressure solution, when time discretization and material properties are chosen in the relevant range to model the conducted field experiments. Hence, this work refrains from treating the matrix by Biot's full theory but approximates the material behavior with Gassmann's low frequency result \cite{gassmann1951,mavko2009}.

The intact gneiss exhibits a permeability $<10^{-20}$ m$^2$ \cite{adero2020}. Thus, leak-off from a fracture, into which fluid is injected from a borehole, into the "surrounding" is controlled by its intersection with other fractures. The hydraulic testing in BH10 revealed that the pre-existing fractures in the gneiss exhibit vastly variable hydraulic properties, as for example evidenced by the results of the pressure-pulse tests (Table \ref{tab:interval_props}). We thus face a range of possible scenarios for the induced or pre-existing fractures intersecting the borehole. They may intersect only poorly permeable pre-existing fractures or linking up with a highly permeable pre-existing fracture. We consider either scenario to be suitable for an approximate description that neglects leak-off. For the second scenario, our modeling will simply gain the equivalent properties of a single fracture, since a variation of properties along a fracture is not tackled and thus a distinction of "individual" fractures composing a conduits is not possible. Neglecting leak-off likely overestimates "effective" length because all of the injected fluid volume has to be stored in the fracture. Applying a single fracture model with fixed geometry to this mix of fractures intends to test the versatility of the model and to determine equivalent fracture properties in a consistent way.     

\subsection{Constitutive Relations}
Traditionally, normal contact models are expressed in terms of fracture deformation relative to the position corresponding to the first, stress-free contact of the two fracture surfaces, where fracture closing is described by positive deformation values \cite{bandis1983,gens1990}. For the response of the fracture to changes in normal stress, we use a modified non-linear elastic constitutive relation based on the model proposed by \cite{gens1990,segura2008} 
\begin{equation}
    \label{eq:normal_deformation_stress}
        \sigma^{\text{FR}}_{\text{N}} = E^\text{Fr} \frac{U^e}{U^\text{max} - U^e}
\end{equation}
that characterizes the normal deformation $U^e$ of an interface with two parameters, the initial stiffness at vanishing normal stress, $E^{\text{Fr}}_\text{eq}$, and the maximum displacement for infinite stress $U^\text{max}$. To be consistent with the governing flow eq.~(\ref{eq:governing}), we formulate (\ref{eq:normal_deformation_stress}) in terms of relative aperture changes
\begin{align}
    \label{eq:normal_deformation_relations}
    \begin{split}
        U^e &= -(\delta - \delta^\text{mech}_0) = -\Delta \delta, \\
        U^\text{max} &= -\Delta \delta_\text{max} = -(\delta_\text{min} - \delta^\text{mech}_0)
    \end{split}
\end{align}
where the fracture deformation $U^e$ is defined as changes of the hydraulic aperture $\delta$ relative to the initial mechanical aperture $\delta^\text{mech}_0$ and the maximal deformation $U^\text{max}$ is defined with respect to the difference between the minimal mechanical fracture aperture $\delta_\text{min}$, approached for infinite normal stress, and the initial aperture, obviously obeying $\delta^\text{mech}_0>\delta_\text{min}$. Neither absolute values of nor changes in mechanical and hydraulic apertures of fractures do have to coincide; particularly true once contact is established and the effective values of these aperture measures strongly depend on contact details and percolation characteristics in the fracture plane \cite{pyraknolte1988}. The resulting hydraulic, respectively mechanical fracture aperture might then be expressed in terms of 
\begin{align}
    \label{eq:normal_aperture_stress_2}
    \begin{split}
    \delta_\text{hyd} &= \delta_0+\Delta \delta, \\
    \delta_\text{mech} &= \frac{\delta_0}{s_0}+\Delta \delta - \delta_\text{min}.
    \end{split}    
\end{align}
where we simplistically relate initial mechanical and initial hydraulic fracture apertures by  $\delta^\text{mech}_0=\delta_0/s_0$. Introducing the constant, dimensionless parameter $s_0\ge1$ intends to distinguish pore space accessible for fluid flow characterized by the hydraulic aperture and the mechanical response of the contact surface characterized by the mechanical aperture \cite{pyraknolte1988,pyraknolte2000,pyraknolte2016}. Inserting (\ref{eq:normal_deformation_relations}) into (\ref{eq:normal_deformation_stress}) gives
\begin{equation}
    \label{eq:normal_aperture_stress}
    \sigma^{\text{Fr}}_\text{N} = -E^{\text{Fr}} \frac{\Delta\delta}{(\frac{\delta_0}{s_0}+\Delta \delta) - \delta_\text{min}}\quad.
\end{equation}

In principle, coupled hydro-mechanical simulations of deformable fractures require to numerically determine the equilibrium state of a fracture before the perturbation of its mechanical state, here associated with the pumping operations. Instead, we reformulate (\ref{eq:normal_aperture_stress}) using the aperture $\delta_\text{eq}$ that reflects the unperturbed in-situ normal stresses $\sigma^{\text{Fr}}_\text{N,eq}$: 
\begin{align}
    \label{eq:normal_in_situ_stress2}
    \begin{split}
    \Delta \sigma^{\text{Fr}}_\text{N} &= \sigma^{\text{Fr}}_\text{N}-\sigma^{\text{Fr}}_\text{N,eq} \\
    &= -E^{\text{Fr}}\left[ \frac{\Delta\delta}{(\frac{\delta_0}{s_0}+\Delta \delta) - \delta_\text{min}} - \frac{(\delta_\text{eq}-\frac{\delta_0}{s_0})}{(\frac{\delta_0}{s_0}+(\delta_\text{eq}-\frac{\delta_0}{s_0}) - \delta_\text{min}}\right]
    \end{split}\quad,
\end{align}
i.e., we shift the reference state of the fracture to the in-situ stress level. Simple manipulations yield the relation in its implemented form
\begin{align}
    \label{eq:normal_in_situ_stress}
    \begin{split}
    \Delta \sigma^{\text{Fr}}_\text{N} &= -\left[E^{\text{Fr}} \frac{\frac{\delta_0}{s_0}-\delta_\text{min}}{\delta_\text{eq}-\delta_\text{min}}\right] \frac{\Delta\delta_\text{eq}}{(\frac{\delta_0}{s_0}+\Delta \delta_\text{eq}) - \delta_\text{min}}\\
    &= -E^{\text{Fr}}_\text{eq} \frac{\Delta\delta_\text{eq}}{(\frac{\delta_0}{s_0}+\Delta \delta_\text{eq}) - \delta_\text{min}}
    \end{split}\quad,
\end{align}
where $\Delta \delta_\text{eq} = \delta - \delta_\text{eq}$ defines the relative aperture change regarding the equilibrium fracture aperture and $E^{\text{Fr}}_\text{eq}$ is introduced as the normal stiffness parameter of the equilibrium state.

The reduction of the numerical model to a single fracture interacting with the low permeable surrounding gneiss resulting in negligible exchange between fracture and solid domain similar to undrained conditions, negligible contribution of shear forces due to the low viscosity of the pore fluid, water, and by the low frequency range ($\ll \,100\,$Hz) of the perturbations induced by the step-rate tests motivates the treatment of the surrounding matrix by a single phase formulation and neglect of the leak-off term $\MakeUppercase{\romannumeral 4{)}}$. Hence, the deformation state of the linear-elastic rock matrix, embedding the fracture, is characterized by effective bulk $K_{eff}$ and shear $\mu_{eff}$ modulus \begin{align}
    \label{eq:gassmann}
    \begin{split}
    K_\mathrm{eff}&=\frac{\phi_0 \left(\frac{1}{K^\mathfrak{s}}-\frac{1}{K^\mathfrak{f}}\right) + \frac{1}{K^\mathfrak{s}} -\frac{1}{K}}{\frac{\phi_0}{K}\left(\frac{1}{K^\mathfrak{s}}-\frac{1}{K^\mathfrak{f}}\right) + \frac{1}{K^\mathfrak{s}}\left( \frac{1}{K^\mathfrak{s}} - \frac{1}{K} \right)} \\
    \mu_\mathrm{eff} &= \mu
    \end{split}
\end{align}
representing Gassmann's low frequency result \cite{gassmann1951,mavko2009}. In eq.~(\ref{eq:gassmann}), $\phi_0$ denotes the initial porosity of the porous matrix, $K^\mathfrak{s}$ the (average) modulus of the compressible grains composing the matrix, $K$ and $\mu$ the bulk and the shear modulus of the dry skeleton, and $K^\mathfrak{f}$ the bulk modulus of the fluid.  

\subsection{Numerical Model}
The flow model requires input values for the elastic parameters of the matrix $K$ and $\mu$, the fluid bulk modulus $K^\mathfrak{f}$, the initial fracture opening $\delta_0$, the equilibrium-normal stiffness parameter $E^{\text{Fr}}_\text{eq}$, the dimensionless contact parameter $s_0$, and flow-boundary conditions for the intersection of the fracture plane with the borehole, as prescribed by the individual experimental protocols followed for the tests in the six intervals. The chosen parameters are listed in Table \ref{tab:material_parameters_fits}; while Freiberg gneiss is anisotropic (see \cite{adero2020}), for simplicity, we rely on representative isotropic material parameters. Freiberg gneiss appears homogeneous in the tested rock volume and we therefore employ a constant value of $4$ for the parameter $s_0$, determined from exploratory calculations. The chosen value represents a rather distinct difference between initial hydraulic and mechanical fracture apertures. The remaining model parameters, the initial hydraulic fracture aperture $\delta_0$ and the equilibrium fracture normal stiffness $E^{\text{Fr}}_\text{eq}$ along with the fracture length $l_\text{Fr}$ resulting from the discretization of the modelled domain, determine the effective hydraulic conductivity and the storage capacity of the tested fractures and the initial normal stress acting on the fracture surfaces. We seek optimized values for these by analysing the misfit between numerical pressure transients and observed pressure transients. 

\renewcommand{\arraystretch}{1.5}
\begin{table*}[htb]
\caption{Matrix and fracture domain parameters used for the numerical fitting of characteristic fracture properties.}
\centering
\begin{adjustbox}{max width=\textwidth}
\begin{tabular}{llllll}\hline
\rule[1.9ex]{0ex}{1ex}\bf{Quantity} & \bf{Value} & \bf{Unit} & \bf{Quantity} & \bf{Value} & \bf{Unit} \\[1.1ex]\hline
\textbf{\textit{Rock parameters}} &&&&&  \\
dry frame bulk modulus $K$ & $2.75 \cdot 10^{1}$ & [GPa] & grain bulk modulus $K^\mathfrak{s}$ & $6.0\cdot 10^{1}$ & [GPa] \\
shear modulus $\mu$ & $1.7 \cdot 10^{1}$ & [GPa] & initial porosity $\phi_0$ & $1.0\cdot10^{-2}$ & [-]  \\
fluid compressibility $\beta^f$  & $4.17 \cdot 10^{-1}$  & [1/GPa] &
effective bulk modulus $K_{eff}$ & $4.25 \cdot 10^{1}$  & [GPa] \\  effective shear modulus $\mu_{eff}$  
& $1.7 \cdot 10^{1}$  & [GPa] &&& \\
\textbf{\textit{Fracture parameters}} &&&&&  \\
contact characteristic $s_0$ & $4.0$  & [-] &  
fluid compressibility $\beta^f$  & $4.17 \cdot 10^{-1}$  & [1/GPa] \\
\hline
\end{tabular}
\end{adjustbox}
\label{tab:material_parameters}
\end{table*}
\renewcommand{\arraystretch}{1.0}
The assumption of a radial-symmetric fracture geometry and a linear-elastic response of the poro-elastic matrix reduce the total number of degrees of freedom (DoF) due to a reduction of the dimension and the neglect of pore-pressure effects in the surrounding domain, respectively. The reduction of DoF results in a high efficiency of the method; simulations of transients require just several minutes on a standard desktop PC with the used numerical discretization corresponding to around $20.000$ DoF for the whole set of modelled fractures.

\subsection{Quantification of misfit}
Identifying "the" best numerical fit requires examination of the evolution of the misfit between observed transients $\boldsymbol{p}_\mathrm{exp}$ and the modeled transients $\boldsymbol{p}_\mathrm{num}$. We quantify misfit by a normalized $L_2$-error measure
\begin{equation}
\label{eq_res:error}
e_{L_2} = \frac{\left\|\boldsymbol{p}_\mathrm{num}-\boldsymbol{p}_\mathrm{exp} \right\|_2}{\left\|\boldsymbol{p}_\mathrm{exp}\right\|_2},
\end{equation}
i.e., misfit is reduced to a single scalar value for each considered parameter combination. The variation of the misfit with combinations of the model parameters, i.e., the existence of global and/or local minima, is not known a priori, but is required to determine the quality and variance of best numerical fits. Iso-surfaces of misfit in the three-dimensional space of the model parameters $\{\delta_0, E^{\text{Fr}}_\text{eq}, l_\text{Fr}\}$ were gained from interpolation between the discrete values of actually performed calculations.  

\subsection{Strategy of parameter search}
We separately investigated the misfit spaces for the pressure-plateau group and the positive-tangent group. The sensitivity of the model to its parameters was studied in a total of $1144$ and $735$ simulations for the pressure-plateau and the positive-tangent group, respectively. Numerical fits suitable for characterization of tested fractures should possess a normalized error of approximately $e_{L_2}\le0.05$ corresponding to absolute pressure deviations continuously below $0.2$ MPa, subsuming the effect of flow-rate fluctuations due to irregularities of the pump, and intrinsic accuracy of sensors and the correction of flow rate for storage capacity of the injection system. The investigated ranges of the individual parameters (Table \ref{tab:parameter_study}) were defined based on an exploratory analysis starting from educated guesses for the in-situ tests. This exploratory search indicated the existence of a potential misfit minimum below the defined error threshold whose location we then investigated further.
For the subsequent analysis of the remaining sets of measurement data, knowledge about the existence of a local minimum motivated a strategy based on biased guesses to numerically determine parameter combinations resulting in the investigated error ranges corresponding to high quality fits. On average, this resulted in approximately 20-25 simulations per transient.  
\renewcommand{\arraystretch}{1.5}
\begin{table*}[htb]
\caption{Parameter ranges for the studies of the model sensitivity for the pressure-plateau and pressure-tangent group.}
\centering
\begin{adjustbox}{max width=\textwidth}
\begin{tabular}{llllllllll}\hline
&\multicolumn{3}{c}{$E^{\text{Fr}}_\text{eq}$ [MPa]}&
\multicolumn{3}{c}{$\delta_0$ [\SI{}{\micro\metre}]}&
\multicolumn{3}{c}{$l_\text{Fr}$  [m]}\\
\bf{Group}&\bf{min}&\bf{max}&\bf{Inc}&\bf{min}&\bf{max}&\bf{Inc}&\bf{min}&\bf{max}&\bf{Inc}\\
\hline
pressure-plateau group& $4.0$ & $6.4$ & $0.2$ & $30.0$ & $39.0$ & $1.0$ & $20.0$ & $90.0$ & $10.0$ \\
positive-tangent group& $2.4$ & $3.6$ & $0.2$ & $33.0$ & $47.0$ & $1.0$ & $2.4$ & $16.0$ & $2.5$ \\
\hline
\end{tabular}
\end{adjustbox}
\label{tab:parameter_study}
\end{table*}
\renewcommand{\arraystretch}{1.0}

\section{Results}
The parameter study aimed at the identification of parameter combinations yielding good fit between field data and numerical simulations to understand the characteristics of fractures responsible for the two distinct groups of pressure transients. In a first step, we focused on one data set of each group, the transient pressure response $M^p_a$ obtained from hydraulic tests at $51.6$ m borehole depth, representing the pressure-plateau group, and data set $M^t_a$ corresponding to tests conducted at a borehole depth of $24.6$ m, representative for the pressure-tangent group before we used the gained knowledge about the existence and magnitude of a local error minimum to reduce the computational costs of the numerical fitting procedure of the remaining data sets by focusing on parameter combinations resulting in low error values.

\subsection{Parameter study}
\subsubsection{Pressure-plateau group}
\begin{figure}[htb]
\begin{center}
    \resizebox{0.65\textwidth}{!}{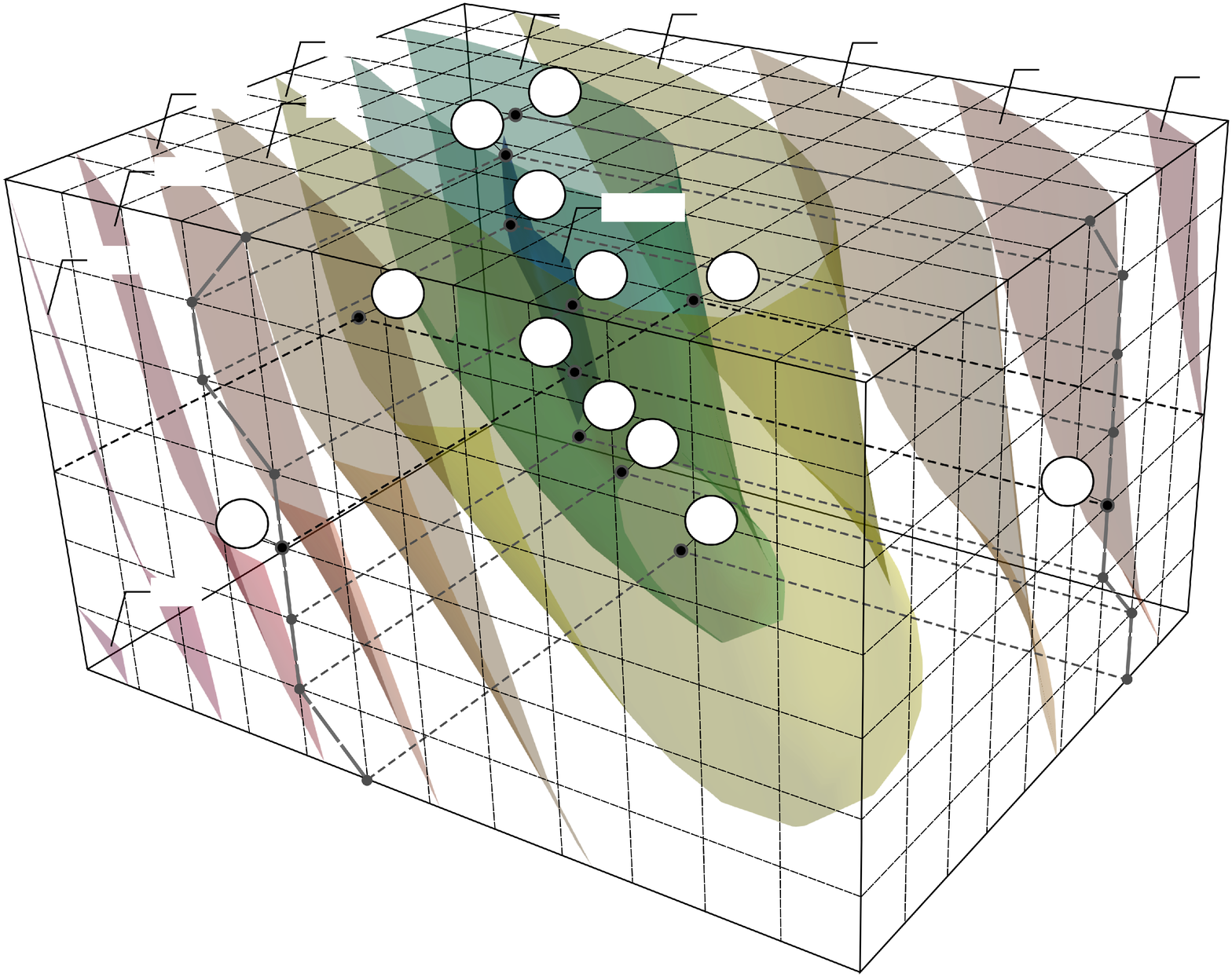} \\
    \resizebox{0.65\textwidth}{!}{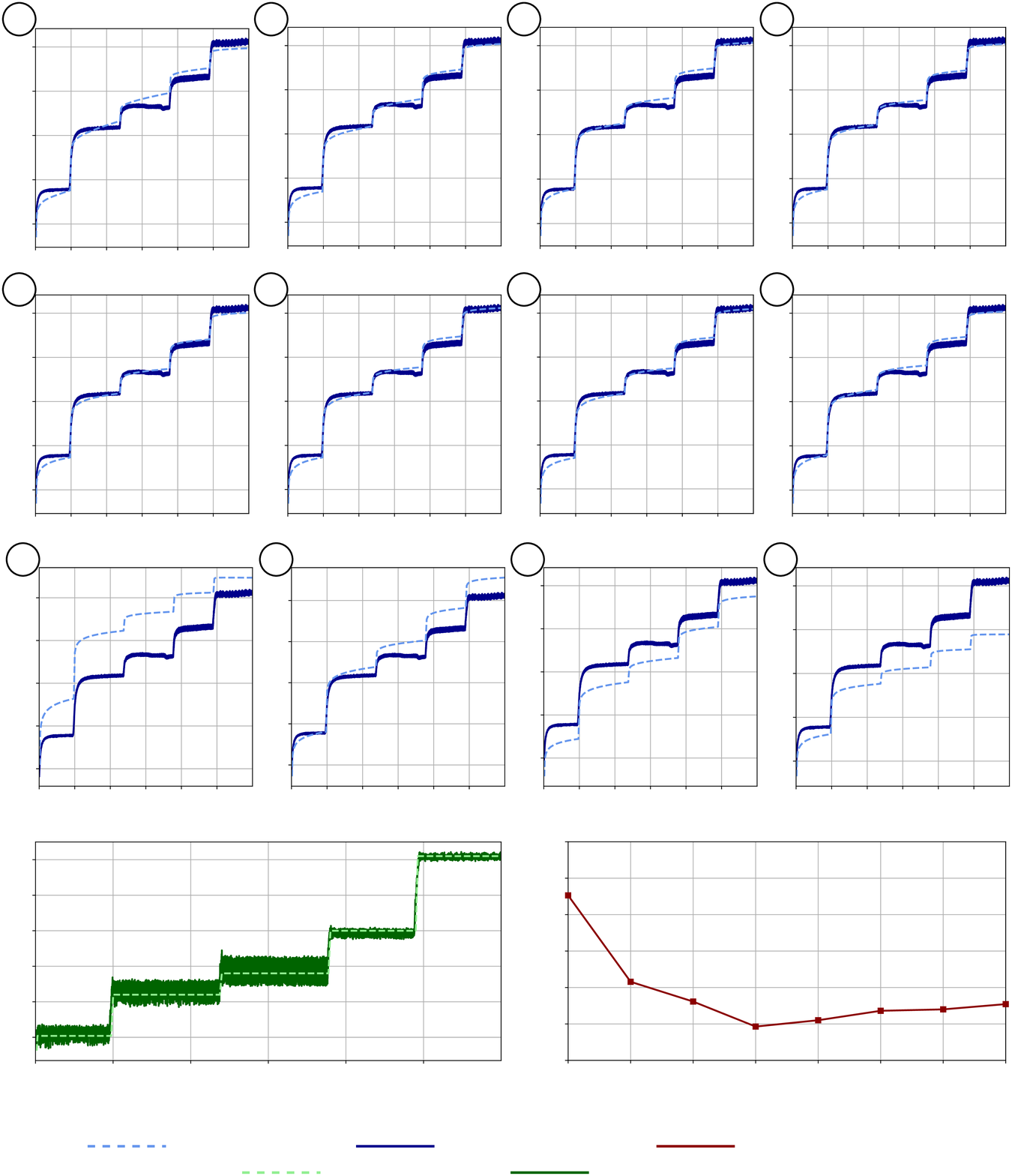}
    \label{fig_res:para_51_6}
    \caption{{\bf Top:} Visualization of iso-surfaces of misfit between observed and calculated pressure transients for the parameter study conducted on the experimental data set $M^p_a$ representative for the pressure-plateau group. Error surfaces are marked with the associated $e_{L_2}$-errors and parameter combinations relevant for detailed analysis of the error evolution are highlighted by labels $A_I$ to $H_I$. {\bf Bottom:} Comparison of observed and numerical pressure transients for the parameter combinations $A_I$ to $H_I$. (left) The flow-rate boundary conditions are well matched by the numerical approach. (right) The errors of the numerical fits for parameter sets $A_I$ to $H_I$ exhibit a minimum. The legend at the bottom applies to all plots.}
\end{center}
\end{figure}

The error surfaces of the pressure-plateau group possess an ellipsoid-like shape (Figure \ref{fig_res:para_51_6}); the closed surface with accurate numerical solutions of $e_{L_2}\le0.0275$ indicates the existence of an error minimum. Examination of the error evolution with fracture length ($A_I$ to $H_I$ in Figure \ref{fig_res:para_51_6}) shows consistency with the introduced iso-surface plot since the error reaches a minimum at a fracture length of $50.0$ m. The error evolution is asymmetric around this minimum, it increases less steep for fractures with an increasing than for fractures with a decreasing length. The corresponding unique error minimum in the parameter space occurs for the parameter combination $D_I$, which consists of the initial hydraulic fracture aperture $\delta_0=$\SI{36.0}{\micro\metre}, an equilibrium fracture normal stiffness parameter of $E^{\text{Fr}}_\text{eq}=5.6$ MPa, and a fracture length of $50.0$ m. The model shows higher sensitivity to the equilibrium-fracture normal stiffness parameter and the initial hydraulic aperture than to fracture length. Individual variations of the fracture stiffness and initial hydraulic aperture relative to the values obtained for the identified minimum (exemplified by $D_I^a$ to $D^d_I$ in Figure \ref{fig_res:para_51_6}) result in pronounced under- and overestimation of the measured transients, respectively. 

\subsubsection{Positive-tangents group}
\begin{figure}
\begin{center}
    \resizebox{0.65\textwidth}{!}{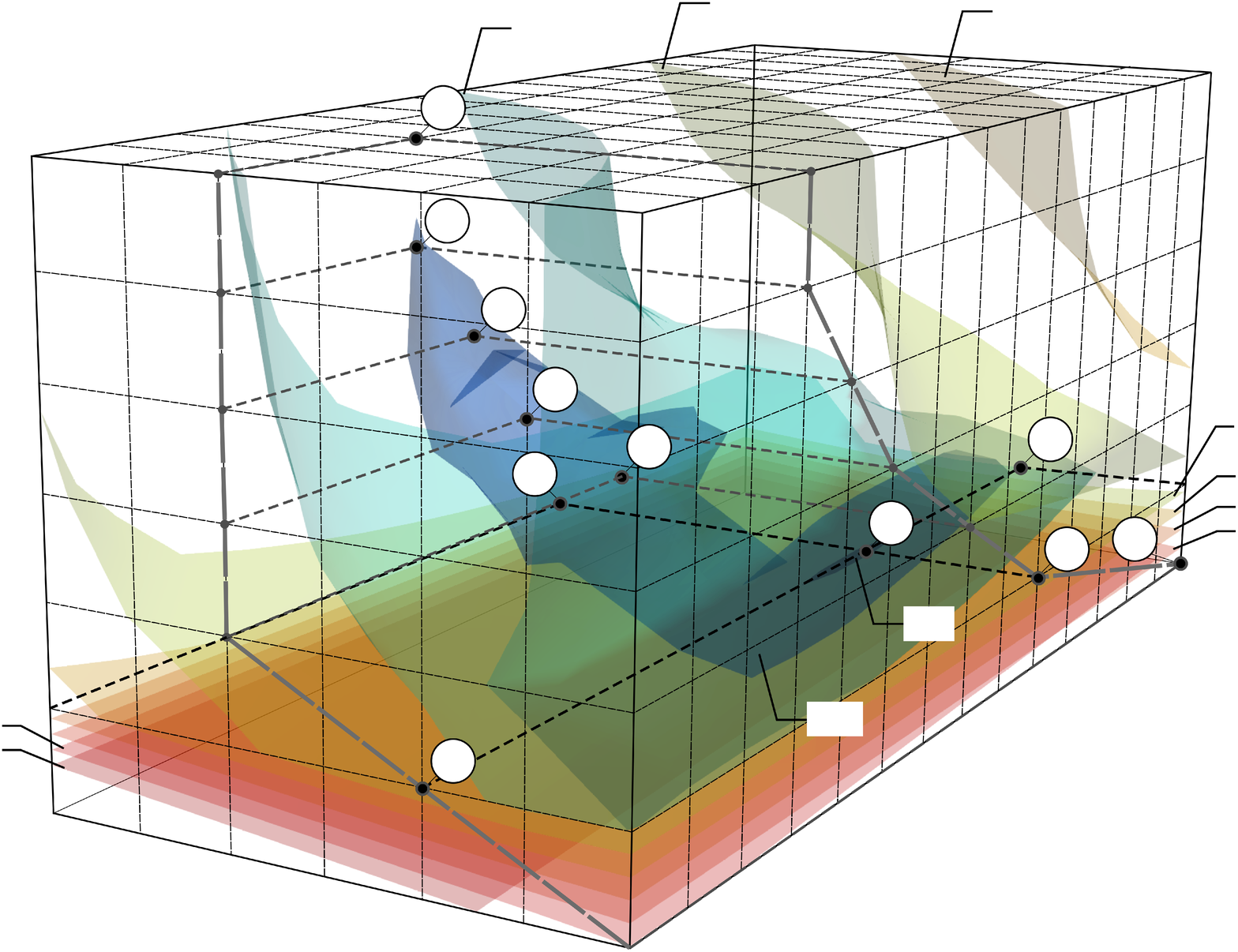} \\
    \resizebox{0.65\textwidth}{!}{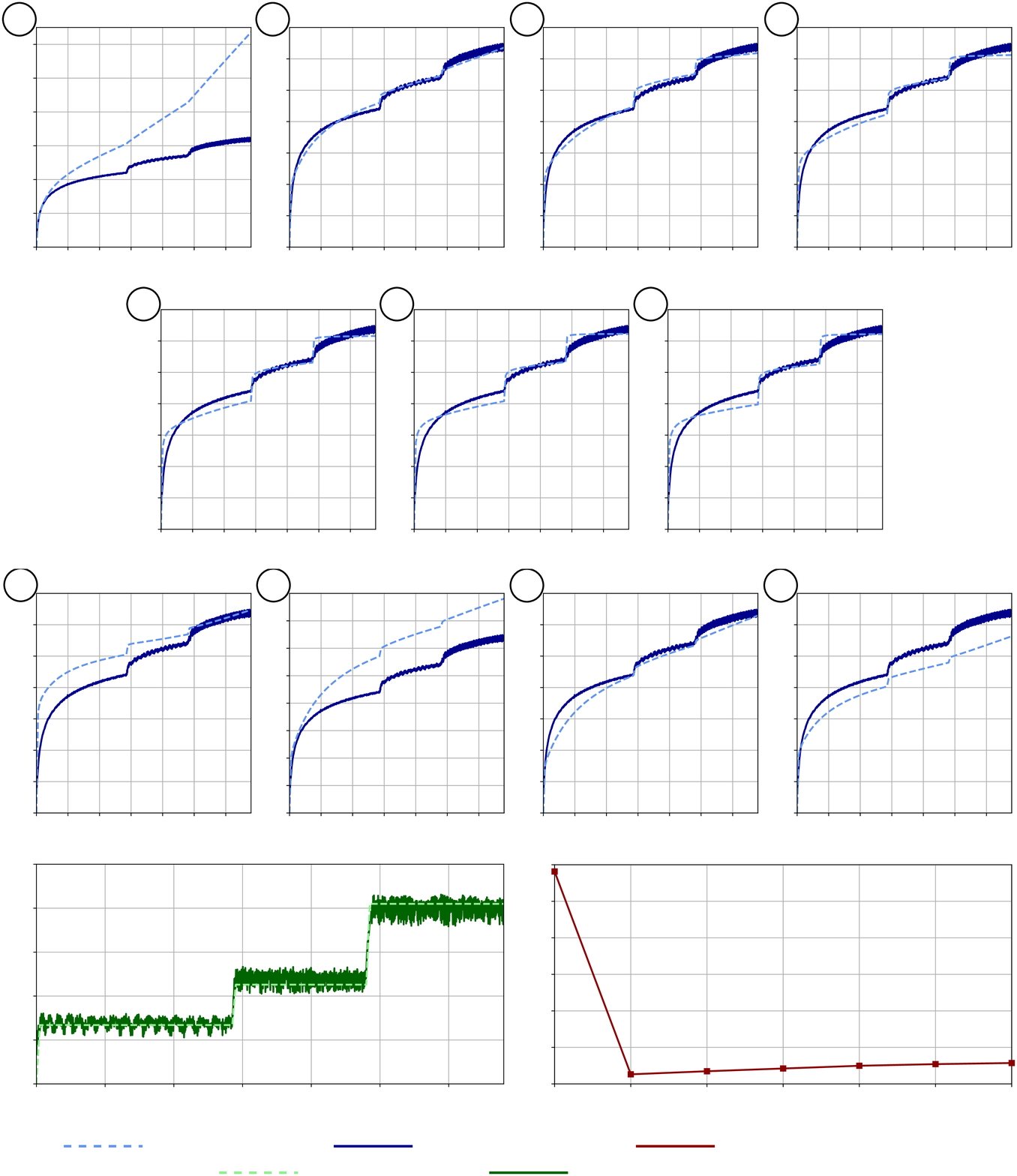}
    \label{fig_res:para_24_6}
    \caption{{\bf Top:} A visualization of error surface plots corresponding to the parameter study conducted on the experimental data set $M^t_a$ representative for the positive group is presented. Error surfaces are marked with the associated $e_{L_2}$-errors and parameter combinations relevant for detailed analysis of the error evolution are highlighted by labels $A_{II}$ to $G_{II}$. {\bf Bottom:} Numerical fits of the measured pressure transients corresponding to the highlighted parameter combinations  $A_{II}$ to $G_{II}$ along with the corresponding fit of flow-rate boundary conditions are introduced. Errors of the numerical fits for parameter sets $A_{II}$ to $G_{II}$ are presented by means of a line plot. A legend introduces the corresponding quantities to the used line types at the bottom of the figure.}
\end{center}
\end{figure}

For the positive-tangent group, the misfit surface identifying relevant parameter combinations with absolute differences to the measured data consistently below $0.2$ MPa, defined by $e_{L_2}\le0.055$, consists of two connected ellipsoidal shapes with different orientations of their axes. A single potential minimum is indicated by the closed iso-surface with $e_{L_2}\le0.03$. The error evolution with fracture length ($A_{II}$ to $G_{II}$ in Figure \ref{fig_res:para_24_6}) confirms the existence of a minimum for the parameter set $B_{II}$, consisting of an equilibrium-fracture normal stiffness parameter $E^{\text{Fr}}_\text{eq}=2.8$ MPa, an initial aperture of \SI{42.0}{\micro\metre}, and a fracture length of $l_\text{Fr} = 4.75$ m. Large misfits result when fracture length decreases below $4.75$ m; however, misfit is less sensitive to variations in fracture length above this value. Variation of the equilibrium-fracture normal stiffness parameter $E^{\text{Fr}}_\text{eq}$ and initial hydraulic aperture $\delta_0$ relative to the parameter set $B_{II}$ (i.e., parameter sets $B^a_{II}$ to $B^d_{II}$ in Figure \ref{fig_res:para_24_6}) reveals a higher sensitivity of the model to changes of the stiffness parameter than the initial hydraulic aperture.

\subsection{Characteristic fracture properties}
\begin{figure}
\begin{center}
    \resizebox{0.75\textwidth}{!}{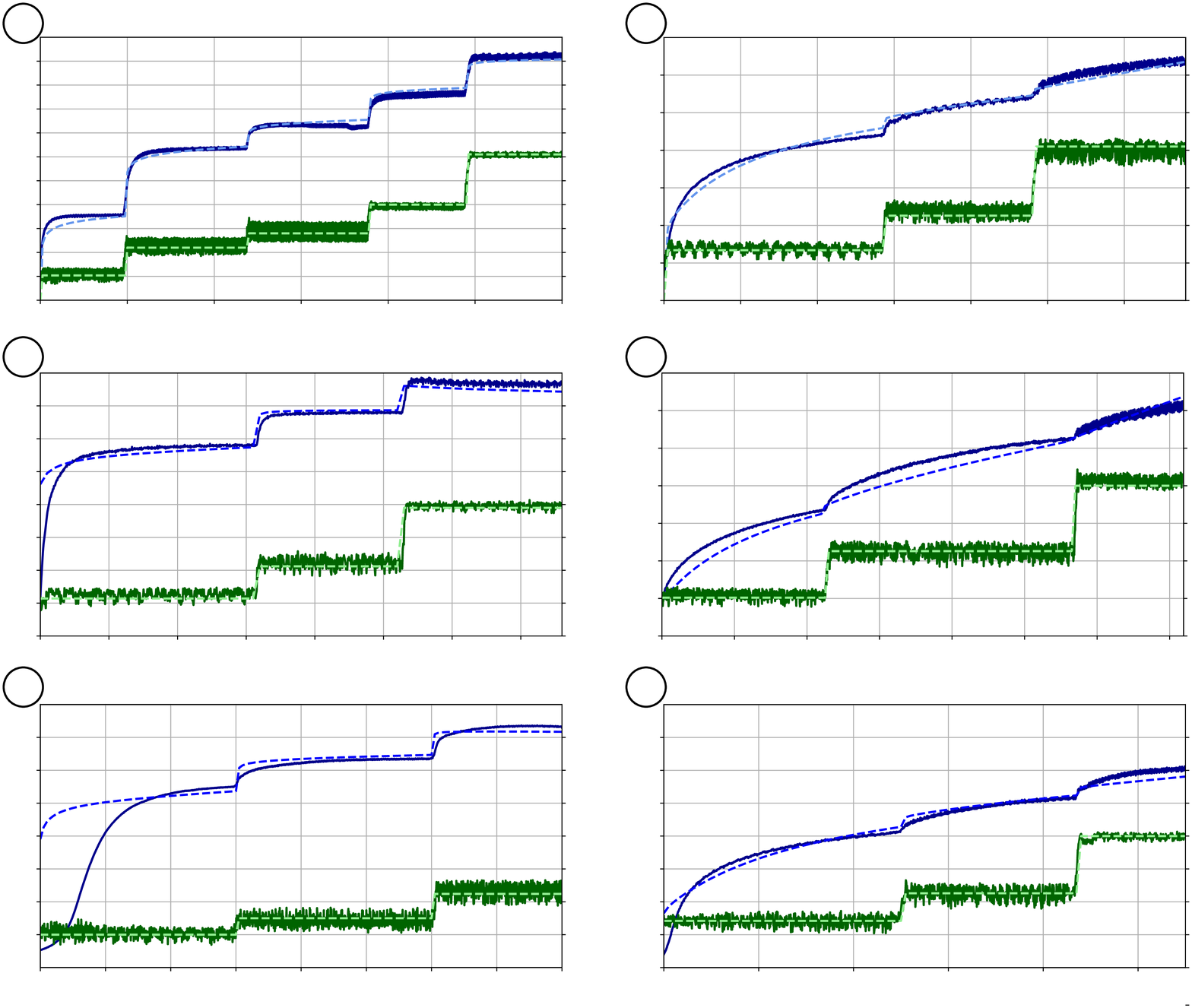} \\
    \resizebox{0.65\textwidth}{!}{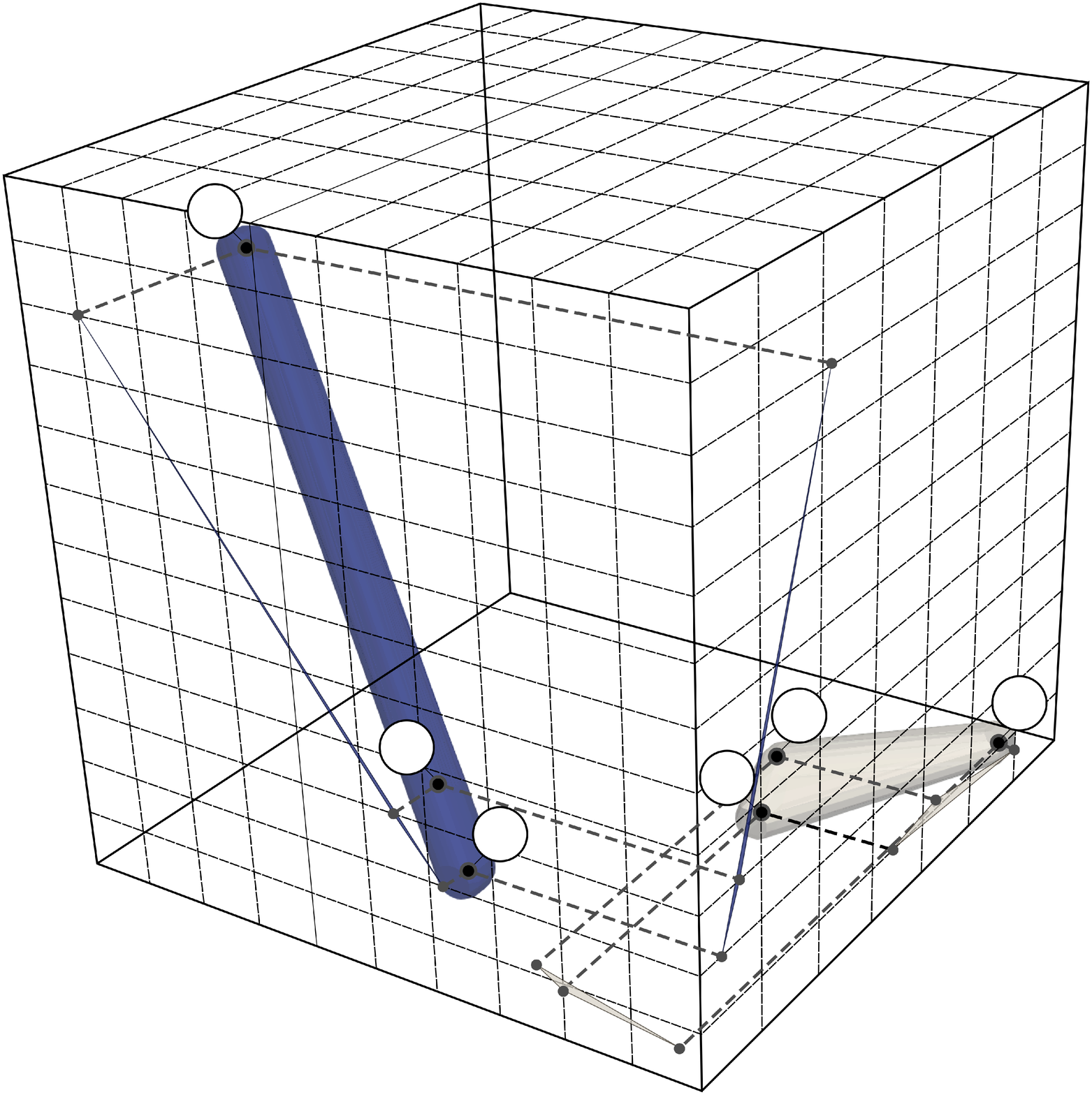}
    \caption{{\bf Top:} Comparison of numerical flow and pressure transients, corresponding to the set of parameter fits determined throughout the numerical fitting, to the experimentally recorded pressure and flow transients. {\bf Bottom:} Visualization of the determined parameter sets in a three dimensional parameter space. The space spanned by parameter combinations associated to the pressure-plateau group is introduced by a blue hull and the space resulting from parameter sets of the positive-tangent group is limited by a grey hull.}
    \label{fig:numerical_fits}
    \end{center}
\end{figure}

For each of the remaining four data sets, parameter sets resulting in fits with error values close to those obtained for the local minima in the two examples above could be determined (Figure \ref{fig:numerical_fits}, Table \ref{fig:numerical_fits}). The error for data set $N^p_c$ is exceptionally high compared to that of other sets when we do not neglect the first pumping step (Table \ref{fig:numerical_fits}) that involves a delayed pressure increase (Figure \ref{fig_res:exp_data}). This interval looses water when isolated and has to be refilled after an extended shut-in period. The identified error minima correspond to distinctly different parameter ranges for the two groups.

The elongated enclosing hull of the determined parameter combinations for the pressure-plateau group visualizes the recognized scaling of fracture stiffness and initial fracture aperture with fracture length. For the positive-tangent group, we observed scaling of the fracture stiffness with fracture length, but no correlation between initial aperture and fracture length does exist (Table \ref{tab:material_parameters_fits}, Figure \ref{fig:numerical_fits}). Optimal parameters of the two groups occupy distinctly different volumes of the misfit space.

\renewcommand{\arraystretch}{1.5}
\begin{table*}[htb]
\caption{Material parameters gained from numerical fitting of the measured pressure transients.}
\centering
\begin{adjustbox}{max width=\textwidth}
\begin{tabular}{llllll}\hline
\rule[1.9ex]{0ex}{1ex}\bf{Depth} &\bf{Fracture label} & \bf{Equ. fracture normal stiffness par.} $E^\text{Fr}$ & \bf{Initial aperture} $\delta_0$ & \bf{fracture length} $l_\text{Fr}$ & \bf{error} $e_{L_2}$  \\[1.1ex]\hline
\textbf{\textit{pressure-plateau group}} &&&&&  \\
$51.6$ m&$N^p_a$ & $5.6$ MPa & \SI{36.0}{\micro\metre} & $50.0$ m & $0.024$\\ %51.6
$55.7$ m&$N^p_b$ & $3.5$ MPa & \SI{26.5}{\micro\metre} & $10.0$ m & $0.048$\\ %49.7
$56.5$ m&$N^p_c$ & $3.8$ MPa & \SI{28.0}{\micro\metre} & $15.0$ m & $0.184/0.024$\\ %56.5
\textbf{\textit{positive-tangent group}} &&&&&  \\
$24.6$ m&$N^t_a$ & $2.8$ MPa & \SI{42.0}{\micro\metre} & $4.75$ m & $0.026$\\ %24.6
$40.6$ m&$N^t_b$ & $2.13$ MPa & \SI{55.0}{\micro\metre} & $3.7$ m & $0.063$\\ %40.6
$49.7$ m&$N^t_c$ & $2.95$ MPa & \SI{46.0}{\micro\metre} & $5.4$ m & $0.046$\\ %55.7
\hline
\end{tabular}
\end{adjustbox}
\label{tab:material_parameters_fits}
\end{table*}
\renewcommand{\arraystretch}{1.0}

\section{Discussion}
Summarizing, throughout the numerical fitting of transient measurement data with pronounced pressure-plateaus a unique minimum in terms of a corresponding parameter combination could be identified in the range of investigated material parameters. The sensitivity analysis proofed continuously increasing errors for relative changes of parameters compared to the set $D_I$ which indicates that no further minima exist in experimentally justifiable limits of the parameters. The model exhibits a high sensitivity for the equilibrium fracture normal stiffness and the initial fracture aperture, whereas simulated pressures are fairly insensitive to changes in fracture length.

\subsection{Capability of the proposed hydro-mechanical model to characterize different pressure transients}
The proposed hydro-mechanical model results in vastly different pressure transients depending on parameter choice and thus either group of observed transients, those with nearly constant pressures and those with continuously increasing pressure at constant flow rate, could be modelled equally well. For both pressure-transient groups, values of initial aperture and equilibrium-normal stiffness parameter determined by the numerical fitting fall well within the range of previous in-situ observations \cite{klimczak2010,schuite2016,zangerl2008}. The fracture lengths of meter-scale derived for the positive-slope group of pressure transients are consistent with the spatial scale of the test volume and the dimensions of seismicity clouds observed during the corresponding stimulations. The decameter-scale fracture lengths modelled for the pressure-plateau group appear long at first glance. Yet, considering the shape of the misfit iso-surface of this group that documents an insensitivity of the model to changes in fracture length beyond a critical lower bound, fracture lengths barely exceeding 10 m cannot be excluded per-se. Furthermore, the model involves only a single fracture, neglecting leak-off into intersecting fracture systems and thus its application to data determines properties of an equivalent fracture potentially subsuming pre-existing fractures with a comparable or higher conductivity than that of the fracture intersecting the borehole.

\subsection{Distribution of pressure along the fracture}
Knowing the distribution of fluid pressure along the fracture is a crucial pre-requisite for substantial stimulation modeling. Since our hydro-mechanical model includes the entire fracture, we can use the determined parameter sets to investigate the pressure distribution in the fracture at any point during the step-rate tests. We focus on the pressure states at the end of each applied flow-rate step for data sets $N^p_a$ and $N^t_b$ that are representative for their corresponding group to examine whether the characteristics of the transients observed in the borehole bear information on the pressure distribution along the fracture. 

\begin{figure}
    \resizebox{1.0\textwidth}{!}{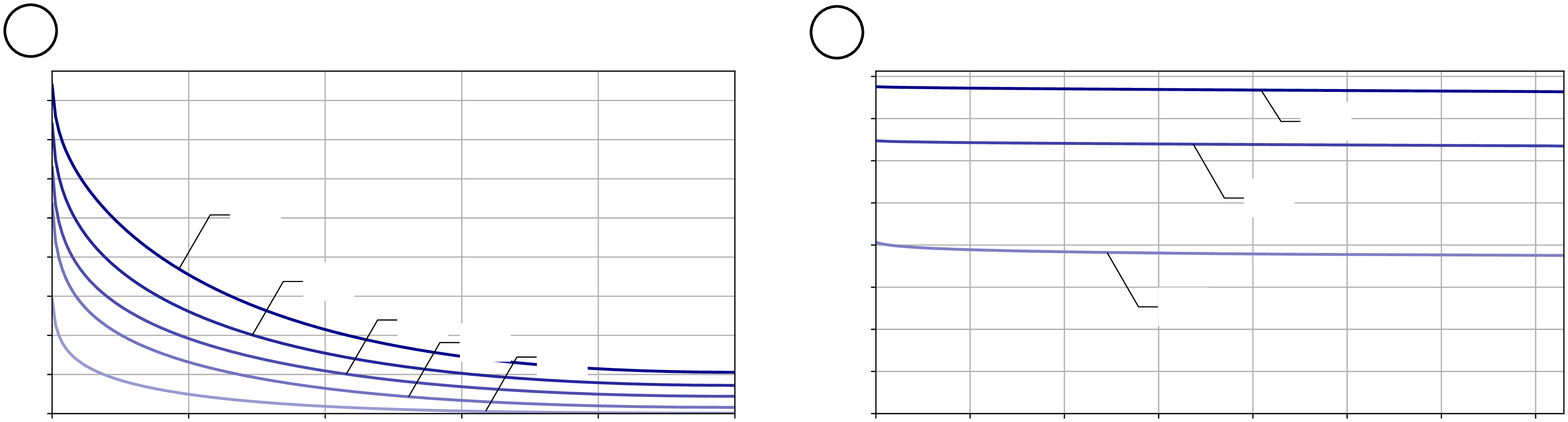}
    \caption{Evolution of pressure distribution along the fractures at the end of each flow-rate step, represented by increasing line thickness with increasing step number $St_i$, for numerical data set $N^p_a$ of the pressure-plateau group (left) and data set $N^t_b$ of the positive-tangent group (right).}
    \label{fig:pressure_evolution}
\end{figure}

For $N^p_a$, the representative of the pressure-plateau group, pressure gradients along the fracture are higher than for $N^t_b$ of the positive-tangent group, for which the pressure profile is almost flat, i.e., the pressure in the fracture is equilibrated during every state of the pumping (Figure \ref{fig:pressure_evolution}). This significant difference in pressure distribution reflects the critical interrelation between local deformation and its consequences for local flow and storage. For the long fractures of the pressure-plateau group, the local deformation and thus permeability decrease with distance from the injection point, but storage of fluid becomes easier close to the borehole where, due to the increased fluid pressure the fracture is already less stiff than at its end. Pressure in the relatively short fractures of the positive-tangent group increases during each flow-rate step, as documented by the continuous increase in pressure in the borehole, and from step to step. The close to constant pressures document that pressure is not controlled by transport restrictions in them but by their storage capacity, which is limited owing to the direct effect of fracture length on fracture volume and also on geometrical fracture stiffness, as detailed in the next section.

\subsection{Specific normal stiffness}
\label{sec:specific_normal_stiff}
\begin{figure}
    \centering
    \resizebox{0.75\textwidth}{!}{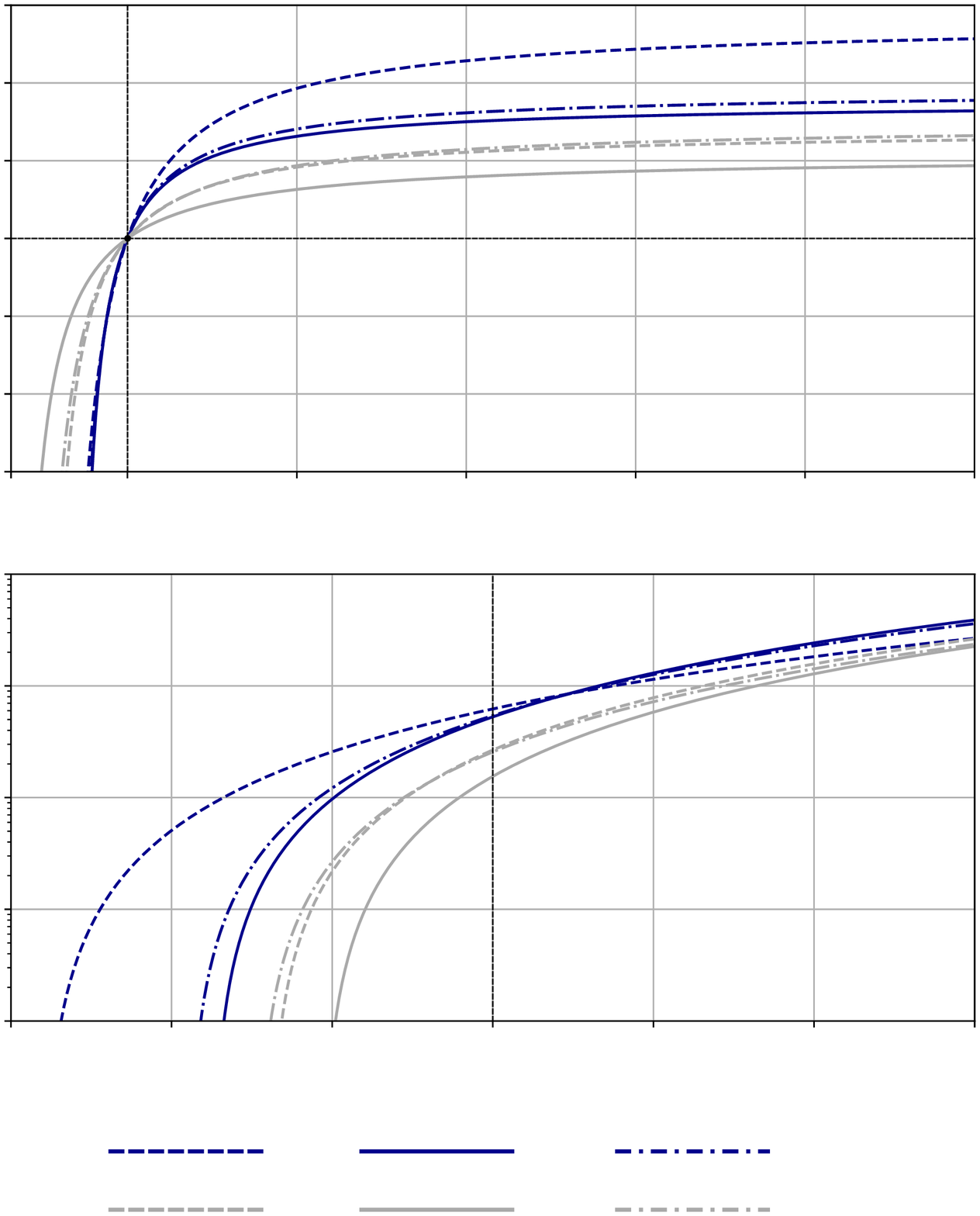}
    \caption{{\bf Top:} Normal contact stress $\sigma^\text{Fr}_\text{N}$ as a function of relative aperture changes $\Delta \delta$ gained from evaluating eq.~(\ref{eq:normal_in_situ_stress}) with the best fit parameters for the investigated pressure transients (Table \ref{tab:material_parameters_fits}). {\bf Bottom:} Semi-logarithmic representation of the specific contact stiffness $K^\text{Fr}_\text{N}$ as a function of the acting normal contact stresses. Positive values indicate compressive stresses (relative to the equilibrium stress). The legend applies to top and bottom, specifically dark blue lines represent data sets of the  pressure-plateau group and grey lines that of the positive-tangent group.}
    \label{fig:stresses_stiffness}
\end{figure}

The contact mechanics of the six investigated fractures is uniquely determined by the parameters constrained by the modeling. Evaluating the constitutive relation (\ref{eq:normal_in_situ_stress}) with the found equilibrium-fracture normal stiffness parameter $E^{\text{Fr}}_\text{eq}$ and the initial hydraulic fracture width $\delta_0$ yields their opening and closure behavior when subjected to normal stresses deviating from the equilibrium stress (Figure \ref{fig:stresses_stiffness}). The corresponding specific contact stiffnesses reflect the strong non-linearity of the constitutive relation; close to the equilibrium stress specific stiffness vary between $10^2$ and $10^3$ MPa/mm and thus fall well within the range of previous in-situ observations and laboratory studies \cite{schuite2016,zangerl2008,pyraknolte2000}. When extended towards fracture closing, i.e., when effective normal stresses exceed the equilibrium in-situ stresses, the stiffness of all tested fractures converge to a rather limited range of $2\cdot 10^3$ MPa/mm to $4\cdot 10^3$ MPa/mm. 

The equilibrium-fracture normal stiffness parameter $E^\text{Fr}_\text{eq}$ defines the tension limit, at which the two fracture halves separate, and thus provides a constraint on the equilibrium stress that the fractures experience in-situ. Since the obtained equilibrium-fracture normal stiffness parameters of the pressure-plateau group are higher than the ones of the positive-tangent group, the predicted equilibrium normal stresses for fractures of the plateau group, ranging between $3.5$ MPa and $5.6$ MPa, are larger than the ones for fractures of the positive-tangent group, ranging from $2.1$ MPa and $2.8$ MPa. Magnitude and range of these predictions are consistent with the stress state inferred for the test volume at Reiche Zeche \cite{adero2020}.

The contribution of the elastic medium, in which the fractures are embedded, to the stiffness of the entire system is conventionally addressed as geometrical stiffness \cite{murdoch2006,vinci2014a}. We evaluated the balance between contact stiffness and geometrical stiffness by prescribing constant fluid pressures in the range of the experimental pressure levels with an increment of $0.5$ MPa for the two equivalent fractures found for intervals $51.6$ m ($M^p_a$) and $40.6$ m ($M^t_b$), representing the two pressure-transient groups and constituting the upper and lower bounds of the parameter space of optimal fits in terms of fracture length and normal stiffness parameter, respectively. 

\begin{figure}
    \centering
    \resizebox{1.0\textwidth}{!}{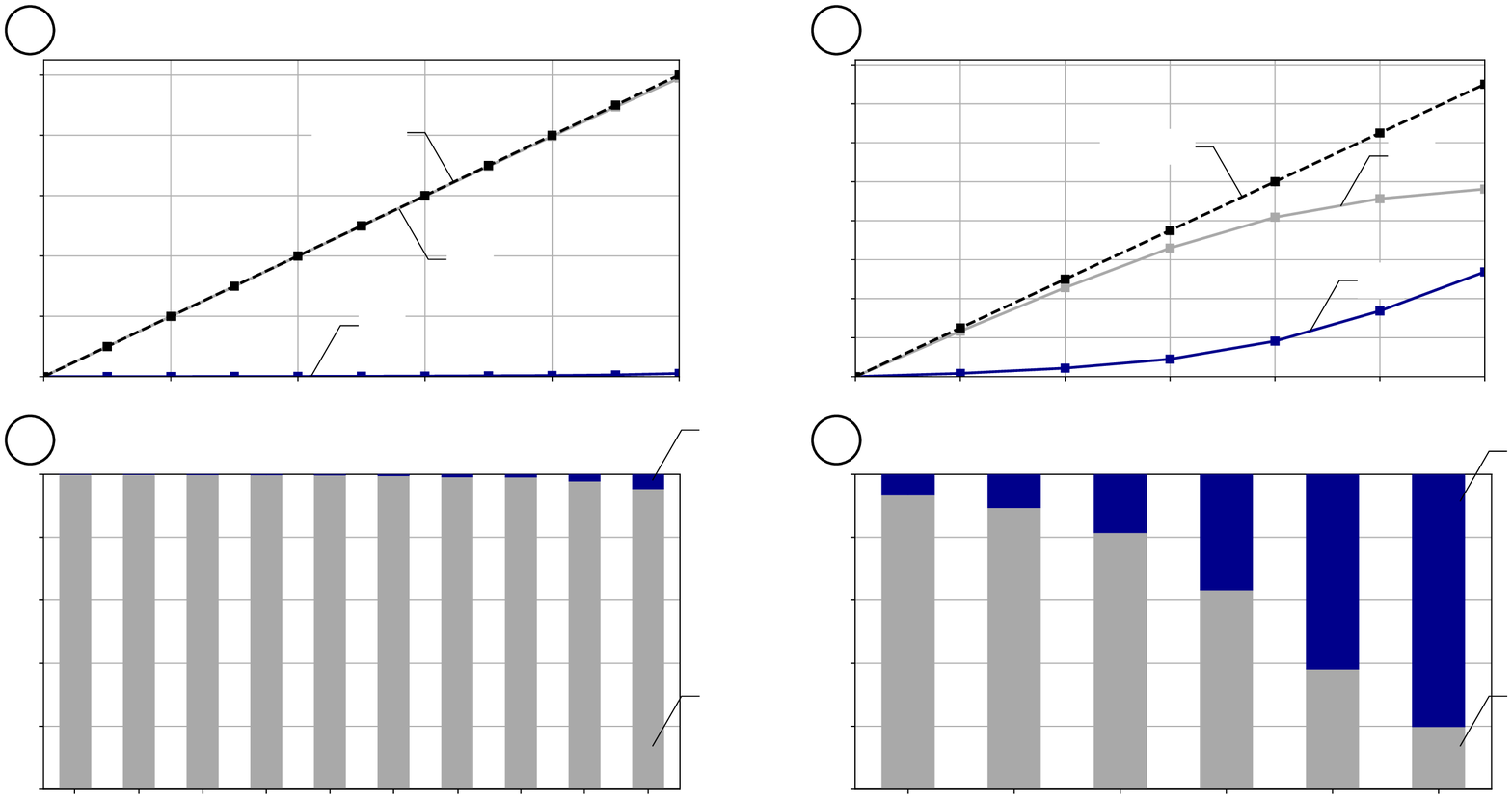}
    \caption{{\bf Top:} Decomposition of the total acting normal stresses into geometrical $\sigma^\text{G}_\text{N}$ and normal contact stress $\sigma^\text{Fr}_\text{N}$ contributions as a function of the acting constant fluid pressure for numerical parameter sets obtained from numerical fits $N^p_a$, representative for the pressure plateau, and $N^t_b$, representative for the positive-tangent group. {\bf Bottom:} Contribution of geometrical and normal contact stresses for each incremental pressure increase.}
    \label{fig:normal_stress_ratios}
\end{figure}

The prescribed levels of fluid pressure lead to local deformations according to the constitutive relation (\ref{eq:normal_in_situ_stress}) and associated local normal contact stresses $\sigma^\mathrm{Fr}_\mathrm{N}$, which we integrate over the fractures' lengths. Mechanical equilibrium across the fracture requires changes in fluid pressure and total normal stress to hold $\Delta p=\Delta \sigma^{\text{Tot}}_\text{N}$. Thus, the mismatch between the applied fluid pressure and the integrated normal contact stresses corresponds to the normal stress exerted on the fracture by the deformation of the surrounding material, here addressed as geometrical normal stress $\sigma^\mathrm{G}_\mathrm{N}$. The decomposition of the changes in total acting normal stress
\begin{equation}
\label{eq_dis:toal_normal_stress}
\Delta \sigma^{\text{Tot}}_\text{N} = \Delta \sigma^{\text{Fr}}_\text{N} + \Delta \sigma^{\text{G}}_\text{N}
\end{equation}
gives changes in the geometrical stress as 
\begin{equation}
\label{eq_dis:geom_normal_stress}
\Delta \sigma^{\text{G}}_\text{N} = \Delta p - \Delta \sigma^{\text{Fr}}_\text{N}.
\end{equation}
The stress balance differs for the two investigated fractures and varies with fluid pressure for an individual fracture (Figure \ref{fig:normal_stress_ratios}). For the long (50 m) fracture of the pressure-plateau group, force balance across the fracture is dominated by contact stresses, while the contribution of geometrical normal stress is significant for the short (3.7 m) fracture of the positive-tangent group, the more the higher the fluid pressure. The changing relative contributions result from the non-linearity of the normal-contact stress formulation introduced by eq.~(\ref{eq:normal_in_situ_stress}). 

We transfer the findings for the normal-stress decomposition (Figure \ref{fig:normal_stress_ratios}) to a corresponding decomposition of specific normal stiffness. Geometrical stiffness of the two investigated fractures is studied by conducting a numerical analysis of their opening behaviour under the assumption of negligible contact normal stresses, which results in $p = \sigma^{\text{G}}_\text{N}$. The discretized geometrical stiffness is then evaluated by means of an averaged aperture and the discretized normal stress change $K^\text{G} = \Delta \sigma^\text{G}_N/\Delta \delta^\text{G}$; fracture contact normal stiffness is calculated by inserting the fitted parameters into the analytical derivative $K^\text{Fr}=\partial \sigma^\text{Fr}_N / \partial \delta^\text{Fr}$. The combined stiffness is numerically determined by $K^\text{Com} = \Delta \sigma^\text{Tot}_N/\Delta \delta^\text{Com}$. Since the relation between $\sigma^\text{Tot}_N=p$, $\sigma^\text{G}_N$, and $\sigma^\text{Fr}_N$ is known from the stress decomposition (Figure \ref{fig:normal_stress_ratios}), the resulting stiffness components can be transformed to express the stiffness of the combined model for a uniform fluid pressure.

\begin{figure}
    \centering
    \resizebox{1.0\textwidth}{!}{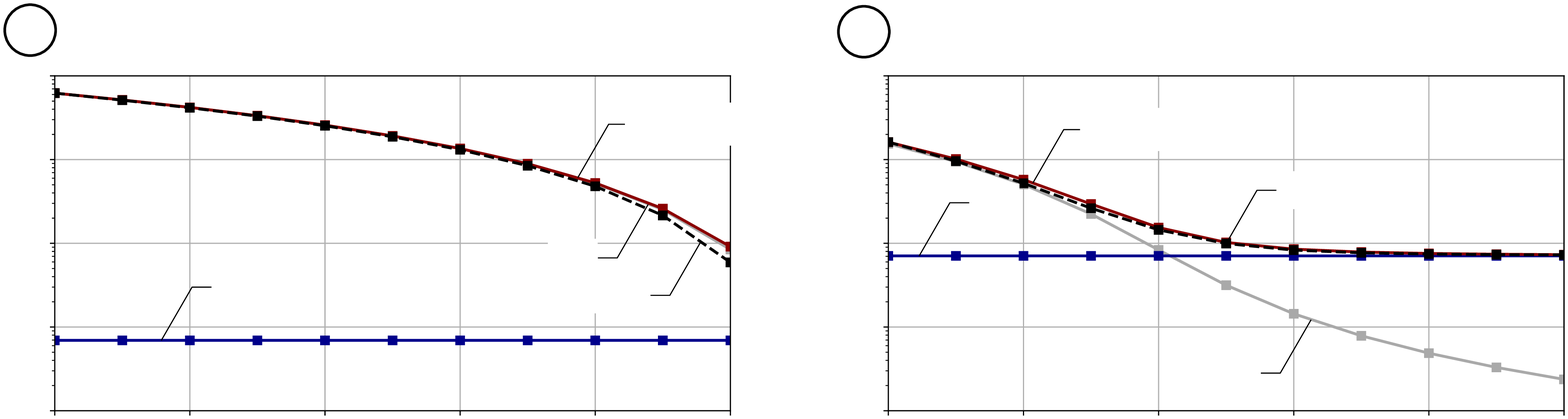}
    \caption{Combined ($K^\text{Com}_\text{N}$) specific fracture stiffness, i.e., the sum of decomposed geometrical $K^\text{G}_\text{N}$ and normal contact stiffness $K^\text{Fr}_\text{N}$, as a function of fluid pressure calculated from results of the proposed hydro-mechanical modelling of data sets $N_a^p$ and $N^t_b$.}
    \label{fig:normal_stiffness_ratios}
\end{figure}

The sum of transformed geometrical and contact normal stiffness agrees with the combined stiffness (Figure \ref{fig:normal_stiffness_ratios}) lending support to the assumptions made regarding the transformation of different stress states to the acting effective normal stress. Geometrical stiffness is found to be negligible for data set $N_p^a$ of the pressure-plateau group, for which the combined specific stiffness is well approximated assuming equivalence of fluid pressure and acting normal contact stresses (see Figure \ref{fig:stresses_stiffness}). In contrast, the combined specific stiffness for data set $N^t_b$, the representative of the positive-tangent group, is a superposition of both stiffness components converging towards the geometrical stiffness with increasing fluid pressure. In fact, the geometrical stiffness forms the lower bound for combined stiffness. The contribution of the geometrical stiffness to the total fracture stiffness is highest for fluid pressures above the identified asymptotic stress values of the contact model, i.e., at the onset of separation of the fracture halves.

\section{Conclusions}
We numerically modeled the opening characteristics of fractures during in-situ hydraulic tests using a hydro-mechanical flow model, here implemented for radial fractures with non-linear contact mechanics and without leak-off. Systematic variations of experimentally observed pressure transients lead us to distinguish two groups, one with continuously (positive-tangent) and step-wise rising (pressure-plateau) pressure response to stepwise increases in flow rate. Our comprehensive parameter study unveiled the effect of fracture length, equilibrium-normal stiffness parameter, and initial fracture aperture on the  the misfit between observed and calculated pressure transients. The proposed hydro-mechanical coupling can explain the strikingly different pressure transients within experimental uncertainty and thus provides a perspective to the response of fractures to pumping operations alternative to the traditional pressure-diffusion analyses, which relate the distinct pressure groups to differences in flow regime related to differences in the orientation of the fracture relative to the borehole 

The identified minima in mismatch between observed and calculated pressure transients correspond to different fracture properties for the two groups, though we noticed a close correlation between fracture length and fracture normal stiffness resulting in a specific mismatch for the positive-tangent group. Pressure plateaus are characteristic of relatively long and stiff fractures, while relatively short and compliant fractures lead to continuously increasing injection pressures. The pressure distribution along the fractures differs significantly for the two groups of pressure transients; pronounced non-linear pressure gradients develop in the long fractures of the pressure-plateau group during injection, while the pressure in the short fractures of the positive-tangent group remains close to the injection pressure along their entire length. Our observations on pressure distribution motivate to investigate the validity of the common practices of normal stress estimation from shut-in and jacking pressures.

Throughout the performed step-rate tests, fluid injection results predominantly in opening of fractures. Yet, the implemented non-linear constitutive relation for the contact mechanics of the fractures covers their opening and closing relative to the initial in-situ state associated with the in-situ stress state. Thus, evaluation of the constitutive relation with the determined fracture parameters allowed us to investigate the contributions of local contacts and overall fracture geometry to stress balance and thus bulk stiffness. With decreasing effective stress, the role of the contacts diminishes and total stiffness approaches the lower bound constituted by the geometrical stiffness.

The proposed hydro-mechanical model exhibits diminished sensitivity to fracture length when a flow-rate step results in a constant injection pressure. Thus, extending the pumping duration will not only help to discriminate between the alternatives of flow regime vs.\ hydro-mechanical effects, but may also reduce uncertainty of model parameters in case pressure ultimately deviates from the apparent early plateau. Future numerical work should explore different scenarios for the relation between mechanical and hydraulic apertures and its evolution with changes in effective normal stresses. 

\section{Acknowledgement/Funding}
The authors gratefully acknowledge the funding provided by the German Federal Ministry of Education and Research (BMBF) for the STIMTEC project (subprojects HYSPALAB and SPATZ, Grant Numbers 03A0015A and 03G0901A) and the GeomInt (I \& II) project (Grant Numbers 03A0004E and 03G0899E), within the BMBF Geoscientific Research Program “Geo:N Geosciences for Sustainability”. Holger Steeb thanks the DFG for supporting this work under Grant No.\ SFB 1313 (Project No.\ 327154368). J\"org Renner is indebted to Felix Becker, Gerd Klee, and Florian Seebald of Solexperts GmbH for the collegial atmosphere during the field testing and Thomas Grelle and Carlos Lehne, and Katja Hesse of LIAG, Hannover, for performing the televiewer logging and processing the data, respectively. 


\end{document}